\documentclass[aps,twocolumn,showpacs,showkeys,superscriptaddress,amsmath,amssymb,nofootinbib,floatfix,prc]{revtex4-1}

\newcommand{\br}[1]{\langle #1\rangle}
\newcommand{\bn}[1]{\left( #1\right)}
\newcommand{\brp}[2]{\langle #1\rangle^#2}
\newcommand{\var}[1]{{\rm var}(#1)}

\usepackage{graphicx}
\usepackage{subfigure}
\usepackage{bm}
\usepackage{hyperref}

\makeatletter

\begin{document}

\title{Multibin correlations in a superposition approach to relativistic heavy-ion collisions}

\author{Adam Olszewski}
\email{Adam.Olszewski.Phys@gmail.com} 
\affiliation{Institute of Physics, Jan Kochanowski University, PL-25406~Kielce, Poland}

\author{Wojciech Broniowski}
\email{Wojciech.Broniowski@ifj.edu.pl} 
\affiliation{Institute of Physics, Jan Kochanowski University, PL-25406~Kielce, Poland}
\affiliation{The H. Niewodnicza\'nski Institute of Nuclear Physics, Polish Academy of Sciences, PL-31342 Krak\'ow, Poland} 

\date{18 February 2015}

\begin{abstract}
Analysis of correlation of multiplicities between various rapidity bins is carried out in the framework 
of a superposition approach consisting of three phases of the ultra-relativistic nuclear 
collision: early partonic phase, intermediate 
collective evolution, and statistical hadronization. 
Simple relations between the moments of produced hadrons and the moments of the sources produced in the initial partonic phase
are presented. They involve only a few effective parameters describing the microscopic dynamics of the system.
We illustrate the practicality of the approach with the Glauber model simulations.
Our study bears direct relevance for the interpretation of the upcoming 
results for the multibin multiplicity correlations from the LHC, which will help in an assessment of the correlation
features of the state formed in the earliest stage of the reaction. 
\end{abstract}



\maketitle

\section{Introduction \label{sec:intro}}

Long-range rapidity correlations between multiplicities of produced 
hadrons have been under active 
consideration since the early experiments  with 
$pp$ and $p\bar{p}$ collisions~\cite{Uhlig:1977dc,Alpgard:1983xp,Ansorge:1988fg,Alexopoulos:1995ft},
followed with heavy-ion data in the RHIC era~\cite{Abelev:2009ag,Tarnowsky:2010qp,ATLAS:2012as}. 
Numerous theoretical investigations followed, in particular%
~\cite{Capella:1978rg,Amelin:1994mf,Braun:2000cc,Braun:2003fn,Brogueira:2007ub,%
Armesto:2007ia,Armesto:2006bv,Vechernin:2007zza,Braun:2007rf,%
Konchakovski:2008cf,Bzdak:2009dr,Lappi:2009vb,Bozek:2010vz,deDeus:2010id,Bzdak:2011nb,Bzdak:2012tp,Vechernin:2012bz,
Bialas:2013xea, De:2013bta,Ma:2014pva}. 
The interest is driven by the expectation that such correlations 
may provide understanding of the elementary space-time dynamics in 
the earliest (partonic) stages of the reaction.
An extension of the conventional forward-backward analysis based on multibin 
correlations in rapidity and applying factorial moments has been 
developed in~\cite{Bialas:2011xk,Bialas:2011vj}.

We recall that according to the interpretation of the STAR Collaboration data of 
Ref.~\cite{Tarnowsky:2010qp,ATLAS:2012as}, made 
in~\cite{Lappi:2009vb,Bzdak:2011nb}, we are quite far from understanding 
theoretically in a complete way the forward-backward correlations in heavy-ion collisions, in particular their 
centrality dependence.
With the analyses from the Large Hadron Collider (LHC) expected to be released in the near future, it is 
important to prepare theoretical ground for the interpretation of the 
data, which is one of the goals of our work.

In this paper, which extends the analysis of Ref.~\cite{Olszewski:2013qwa} to the 
case of three bins (forward, central, and backward), we analyze the 
pseudorapidity correlations in a {\em superposition approach}.
In this framework, to describe the statistical aspects of relativistic heavy-ion dynamics we use a common picture 
involving three separate stages: 1)~early generation of initial partonic densities (sources), 2)~collective (hydrodynamic or transport) evolution, and 
3)~statistical hadronization. A crucial assumption for the method is that the emission from a source is independent from other sources
and universal, i.e., occurs from the same statistical distribution 
independently of the rapidity or the centrality class of the collision.  
This leads to very simple formulas for the statistical measures, relating the moments of the initial sources to the moments of the distribution 
of the produced hadrons, in particular the multibin correlation coefficients.
Parameters present in these relations depend on the features of the overlaid statistical distributions and properties of hydrodynamics, and 
have simple interpretation.

We use the derived relations to obtain predictions for multiplicities, variances, and multibin correlations 
of the produced hadrons. To model the initial phase, we use the Glauber framework based on wounded 
nucleons~\cite{Bialas:1976ed,Bialas:2008zza,Broniowski:2007ft} amended with binary collisions~\cite{Kharzeev:2000ph,SchaffnerBielich:2001qj,Back:2001xy}.
The predictions are made for the Pb+Pb collisions at $\sqrt{s_{NN}}=2.76 \textrm{~TeV}$, as the 
corresponding analyses from the LHC are expected to be published in the near future. 

We note that simulations of the forward-backward multiplicity correlations 
for the Au+Au collisions at the highest RHIC energies were carried out by Konchakovski et al.~\cite{Konchakovski:2008cf} in the framework
of the wounded-nucleon  model with overlaid Poisson distribution and in the Hadron-String-Dynamics transport approach~\cite{Ehehalt:1996uq}. That work 
carefully discusses the influence of centrality selection on the correlation measures used in the experiment, pointing out 
to effects stemming just from the fact that the centrality classes are wide. Our results from the Glauber model comply to these general findings.

Our paper is organized as follows. In Sec.~\ref{sec:thr} we review the three-stage approach, consisting of the early
production, the hydrodynamic evolution, and the final freeze-out. We derive the relations between the multibin moments of the 
sources produced in the early phase and of the multibin moments of final hadrons.
Some general predictions of the approach are given in Sec.~\ref{sec:results}. In Sec.~\ref{sec:cen} we discuss the 
technically important issues 
related to the choice of the centrality bins. For the considered statistical measures, scaling laws with the width of the centrality bin are 
derived. Section~\ref{sec:glis} presents our results of the Glauber simulations for the initial phase. We show here the practicality of our method 
in the application to the future experimental data, which will serve to verify the model of the initial phase.

\section{Three stage approach \label{sec:thr}}

\begin{figure}
\begin{center}
\includegraphics[scale=0.24]{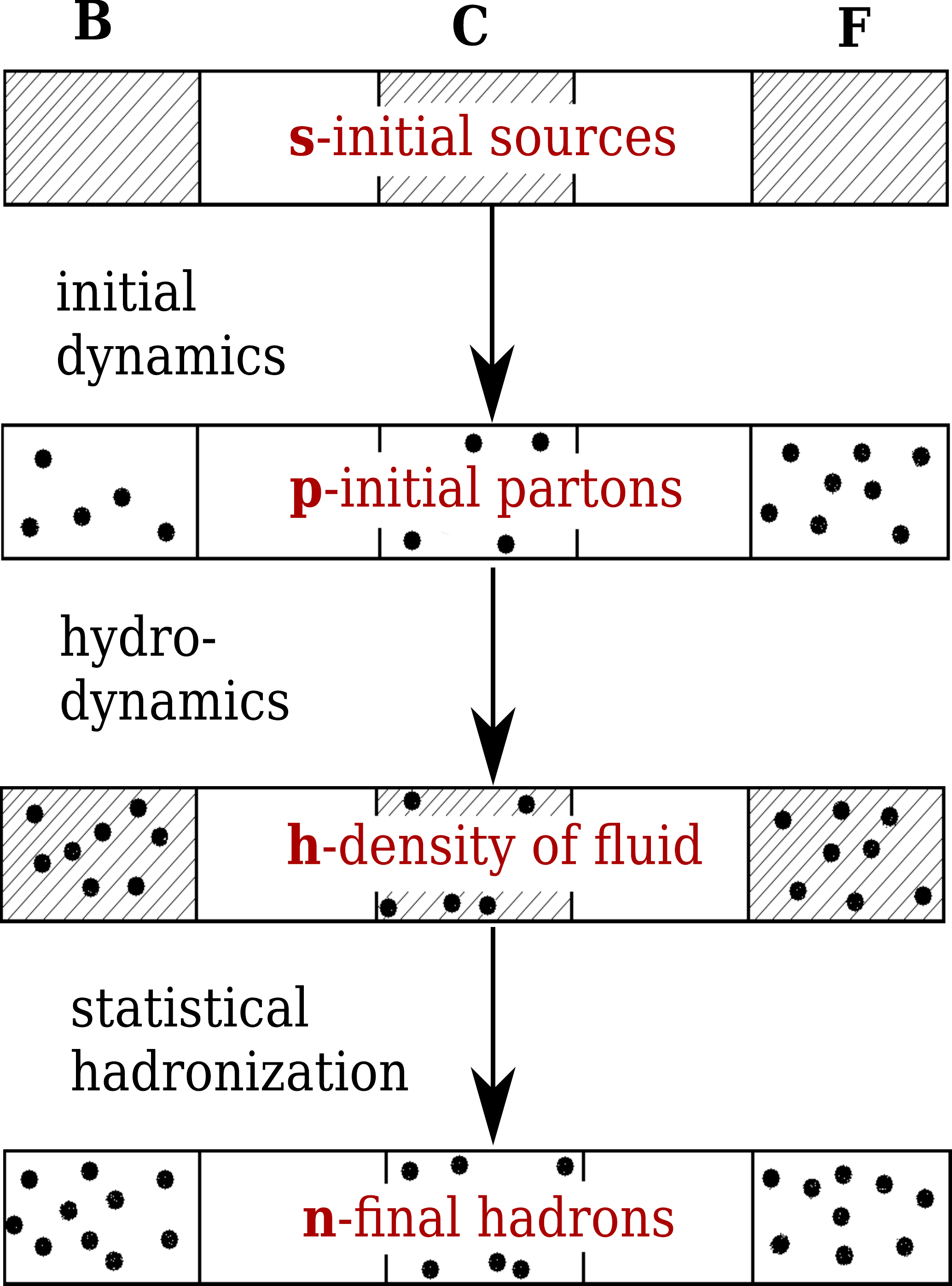}
\caption{(color online) Time evolution of the created system, visualizing the subsequent 
stages of particle production. B, C, and F indicate the backward, central, and 
forward bins, respectively, extending along the spatial rapidity or 
pseudorapidity direction. 
\label{fig:hist}}
\end{center}
\end{figure}

In Ref.~\cite{Olszewski:2013qwa} we have considered a three-stage approach, 
based on superposition, that describes certain features of production of 
particles in relativistic heavy ion collisions and their correlations, focusing entirely on 
statistical aspects. In this framework, the complicated dynamical evolution of 
the system is represented with the following separate stages:

\begin{enumerate}

\item Initial phase, which may be modeled with the Glauber 
approach~\cite{Czyz:1969jg,Bialas:1976ed,Bialas:2008zza,Kharzeev:2000ph,
Back:2001xy,Broniowski:2007ft}, color glass 
condensate~\cite{Kovner:1995ja,Iancu:2000hn,Lappi:2010cp,Lappi:2011zz}, string 
formation~\cite{Amelin:1994mf,Brogueira:2006yk}, etc. As the result of this 
early dynamics, some initial space-time distribution of entropy or 
energy-density is generated. We assume that first the initial {\em sources} are 
formed (see Sec.~\ref{sec:s}), denoted with $s$, which then produce the 
distribution of initial partons, $p$ (see top of Fig.~\ref{fig:hist}). The extent of correlation 
between the source densities in spatial rapidity bins $B$, $C$, or $F$ depends on the 
early dynamics.

\item Hydrodynamic evolution (for
reviews see, e.g.,~\cite{Heinz:2013th,Gale:2013da} and references therein) or 
transport~\cite{Lin:2004en} evolves the initial density of partons 
$p$ into expanding hydrodynamic fluid density $h$ (third row in Fig.~\ref{fig:hist}). 
In our considerations it is only important that $h$ is (to a good approximation) proportional 
to $p$.

\item Freeze-out and the statistical hadronization mechanism (see, 
e.g.,~\cite{WFlorkowski:2010zz} for a review), converting the hydrodynamic 
fluid into hadrons $n$, which are then subject to experimental observation 
in various pseudorapidity bins.
 
\end{enumerate}

The above simple scheme summarizes most of the popular model approaches to 
particle production in relativistic heavy-ion collisions. Our treatment focuses 
on statistical aspects of each of the phases.
While the underlining dynamical mechanisms are very complicated, it turns out 
that certain statistical features may be simply parametrized, and specific 
conclusions concerning particle correlations in rapidity may be, under mild assumptions, drawn 
in a model-independent way.

\begin{figure}
\begin{center}
\includegraphics[scale=0.037]{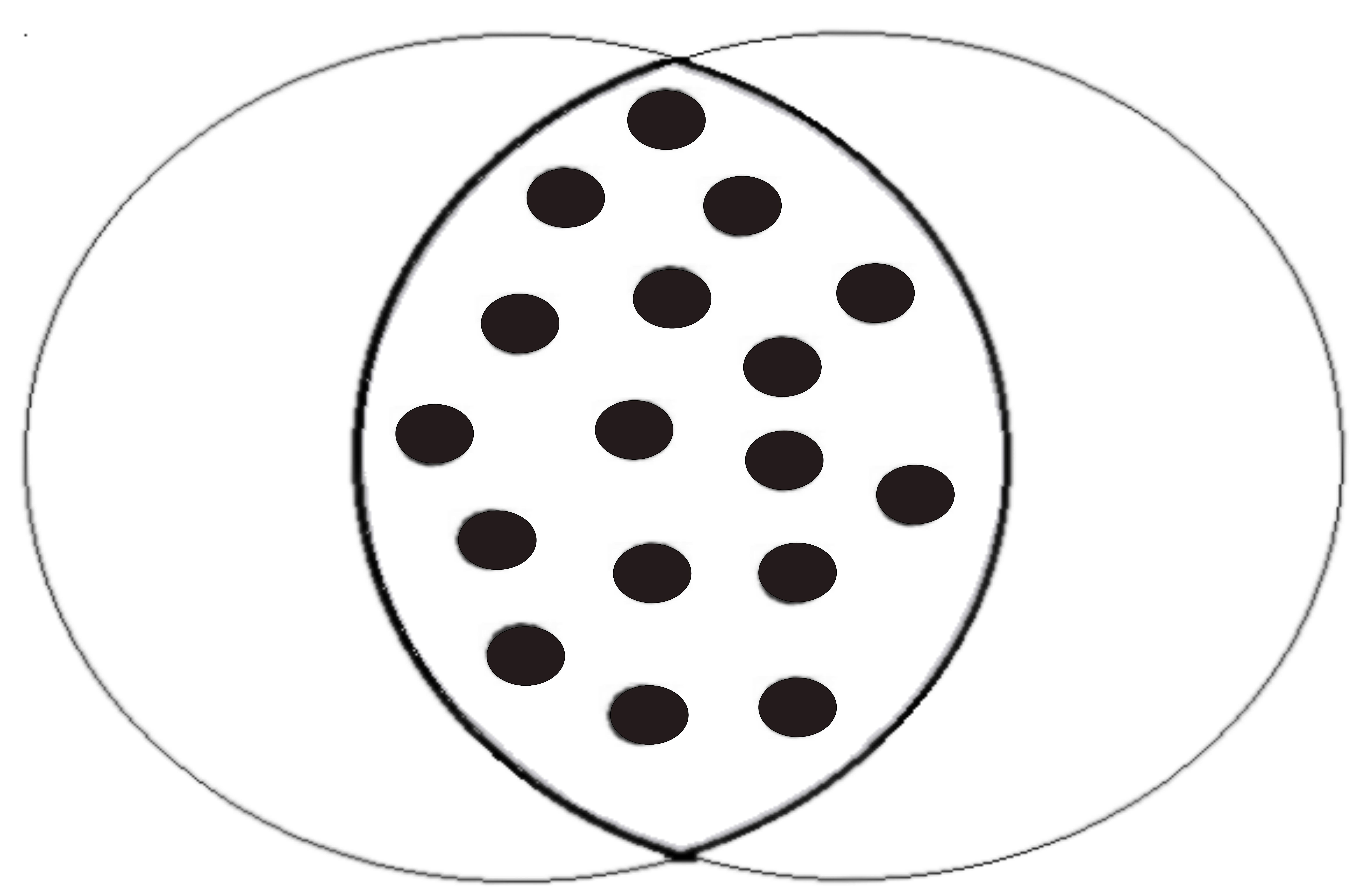}
\includegraphics[scale=0.22]{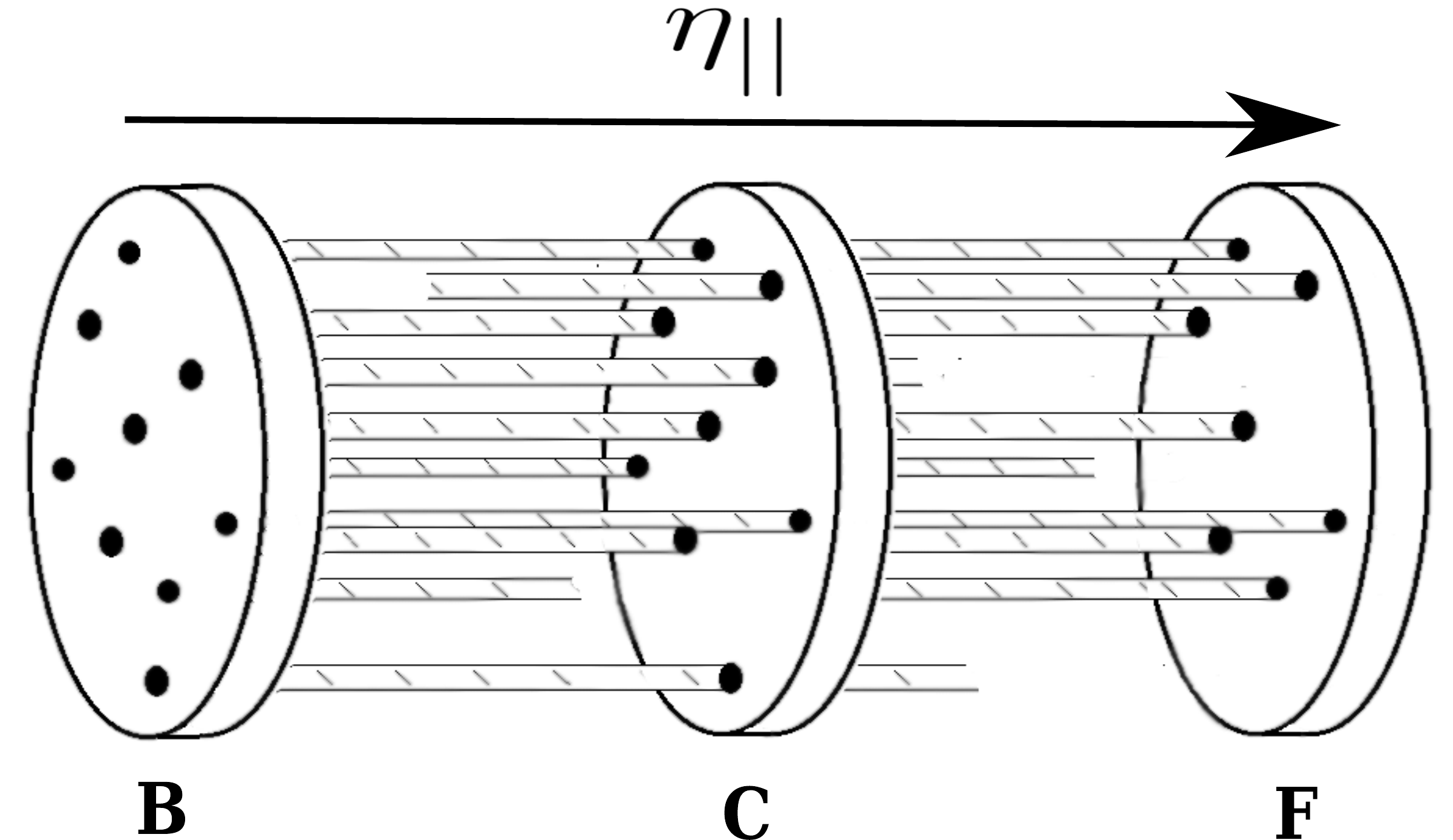}
\caption{(color online) Initial source projected on the transverse plane (left) and the three 
dimensional picture of color flux-tubes extending along the spatial rapidity 
(right). 
\label{fig:nuc}}
\end{center}
\end{figure}

\subsection{Sources \label{sec:s}}

We use the concept of the initial {\em sources}, which yield the initial space-time entropy or energy density distribution. The precise 
definition depends on the model of the initial phase.
In our work we use the Glauber 
model~\cite{RGlauber:1959aa,Czyz:1969jg,Bialas:1976ed,Miller:2007ri,%
Broniowski:2007ft,Bialas:2008zza}, where the source is 
created in an individual NN collision, located in the transverse plane 
and extending along the spatial rapidity direction, 
\mbox{$\eta_\parallel=\frac{1}{2} \log[(t+z)/(t-z)]$} (cf. Fig.~\ref{fig:nuc}). 
A particular model of the distribution in $\eta_\parallel$, coming from
Ref.~\cite{Bozek:2010bi} and used here, is 
described in Appendix~\ref{sec:cenexp}. It is motivated with the success of 
Ref.~\cite{Bialas:2004su} in describing the deuteron-gold collisions~\cite{Nouicer:2004ke}, followed with the 
extension to NN collisions~\cite{Gazdzicki:2005rr,Bzdak:2009dr,Bzdak:2009xq}.
We note that other descriptions of the early production could be used as well, for instance the 
Kharzeev-Levin-Nardi framework
\cite{Kharzeev:2000ph,Kharzeev:2002ei,Drescher:2006pi}.

In certain models, the sources may be interpreted as color flux-tubes, extending along some 
range in rapidity, as pictured in Fig.~\ref{fig:nuc}). It may happen that some 
flux-tubes are longer, extending through the whole region of interest from 
the $B$ to $F$ bins, while some may be shorter, as indicated in 
Fig.~\ref{fig:nuc}), making the correlations shorter-range. Clearly, such effects influence the 
multiplicity correlations between the bins, which is what ultimately we wish to 
investigate. 

The dots in Fig.~\ref{fig:nuc} indicate the sources in bin $A$, where
$A=F,B,C$. Their corresponding numbers are denoted as $s_A$ throughout this 
paper.

\subsection{Centrality definition \label{sec:cendef}}

The number of sources in the central bin, $s_C$, is used to define the centrality 
classes of the collision in our simulations presented in Sec.~ \ref{sec:glis}. In the wounded nucleon model this 
prescription would correspond to using the number of 
wounded nucleons to determine centrality, as is frequently done in model studies. 
The definition based on wounded nucleons or participants of the projectile is appropriate for experiments
which measure the projectile spectators in dedicated detectors (e.g, as in NA61). On the other hand, 
in experiments where the multiplicity of the event is used as the centrality selector, 
a closer measure would be the number of hadrons generated in the model, e.g., $n_C$. 
As pointed out in Ref.~\cite{Konchakovski:2008cf}, the definition of centrality may influence 
the results for the forward-backward correlations to some extent. We investigate this issue in Sec.~\ref{sec:cn}.

\subsection{Initial dynamics}

In the early stage of the reaction, the sources emit partons.\footnote{We 
note that this mechanism was implemented already in the original wounded 
nucleon model~\cite{Bialas:1976ed}, where the Poisson distribution was overlaid 
over the distribution of the wounded nucleons.} We assume that each source 
produces partons independently of one another and, in addition, the 
production occurs independently in the $F$, $B$, and $C$ bins, as they are assumed to be
well separated in the spatial rapidity $\eta_\parallel$. Thus, each source $i$ emits $\mu_i$ partons 
according to the same distribution, 
\begin{eqnarray}
 \mathnormal{  p_A =\sum_{i=1}^{s_A} \mu_i,\quad  A=B,F,C}.
 \end{eqnarray}
It is then elementary to obtain the formulas for the above superposition model 
for moments corresponding to averages over events (here listed up to rank 
three):
\begin{eqnarray}
\langle p_A \rangle &=&  \langle \mu \rangle \langle s_A \rangle, \nonumber \\
\br{\Delta p_A^2} &=& {\rm var}(\mu) \langle s_A \rangle + \langle \mu \rangle^2 \br{\Delta s_A^2},\nonumber  \\ 
\br{\Delta p_A \Delta p_{A'}} &=& \langle \mu \rangle^2 \br{\Delta s_A \Delta s_{A'}},  \nonumber\\
\br{\Delta p^3_A}&=&\mu_3\!\bn{\mu}\br{s_A}+3\var{\mu}\br{\mu}\br{\Delta s_A^2}\nonumber \\
&+&\brp{\mu}{3}\br{\Delta s_A^3},\label{eq:pstat}\\
\br{\Delta p_A \Delta p^2_{A'}} &=& \var{\mu}\br{\mu}\br{\Delta s_A \Delta 
s_{A'}} \nonumber \\ &+& \brp{\mu}{3}\br{\Delta s_A \Delta s^2_{A'}},  
\nonumber\\
\br{\Delta p_F \Delta p_C \Delta p_B} &=& \langle \mu \rangle^3 \br{\Delta s_F \Delta s_C \Delta s_B},  \nonumber
\end{eqnarray}
where $A,A'=F$, $B$, or $C$, and $A\neq A'$. In our notation $\Delta 
p_A^n=(p_A-\langle p_A \rangle)^n$, $n=1,2,3$, while $\langle \cdot \rangle$ 
denotes averaging over events. The symbol $\mu_3(x)$ stands for the third central 
moment of the distribution of $x$.

\subsection{Hydrodynamics}

The density of partons formed in the early phase, as described above, provides 
initial conditions for the equations of the hydrodynamic evolution of the fluid 
cell in the intermediate phase. This evolution, proceeding until freeze-out, is 
assumed to be deterministic (note, however, Refs.~\cite{Kapusta:2011gt,Kapusta:2012gm}), hence no extra fluctuations are introduced at 
this stage. As a result, the initial density of partons $p$ is 
deterministically carried over to the entropy density of the fluid cell at 
freeze-out, denoted with $h$ (cf. Fig.~\ref{fig:hist}). 

In the vicinity of the average value 
of $p$ the function $h(p)$ is approximately affine, hence we may 
expand
\begin{eqnarray}
 h = t_0  \langle p \rangle + t_1 (p - \langle p \rangle ) + {\cal O}\left ( (p - \langle p \rangle )^2 \right ), \label{eq:hs}
\end{eqnarray}
where $t_i$ are parameters depending on the dynamical response of hydrodynamics to 
the initial condition. The higher-order terms in Eq.~(\ref{eq:hs}) may be dropped if $p$ is 
sufficiently close to $\langle p \rangle$, otherwise corrections arise to the 
formulas developed below. With Eq.~(\ref{eq:hs}) we get
\begin{eqnarray}
\langle h_A \rangle &=& t_0  \langle p_A \rangle, \nonumber \\
\br{\Delta h_A^2} &=& t_1^2 \br{\Delta p_A^2},  \nonumber\\
\br{\Delta h_A \Delta h_{A'}} &=& t_1^2 \br{\Delta p_A \Delta p_{A'}}, \nonumber\\
\br{\Delta h_A^3} &=& t_1^3 \br{\Delta p_A^3}, \label{eq:hstat}\\
\br{\Delta h_A \Delta h_{A'}^2 } &=& t_1^3 \br{\Delta p_A \Delta p_{A'}^2 },\nonumber\\
\br{\Delta h_F \Delta h_C \Delta h_B } &=& t_1^3 \br{\Delta p_F \Delta p_C \Delta p_B }.\nonumber
\end{eqnarray}

We note that the approximate linearity of the hydrodynamic response has been 
well determined for the 
evolution of eccentricities~\cite{Gardim:2011xv,Niemi:2012aj} at a given 
centrality. For the present case of multiplicities, the success of modeling of 
the measured particle multiplicity distributions within the Glauber 
framework~\cite{Back:2001xy,Back:2004dy} indirectly supports the assumptions leading to 
Eq.~(\ref{eq:hstat}).

\subsection{Statistical hadronization}

When hydrodynamic evolution ends, statistical hadronization phase takes over at 
freeze-out (cf. Fig.~\ref{fig:hist}). We assume that the well-separated fluid 
cells $h$ emit $n$ hadrons independently from one other into corresponding well-separated
phase-space regions in pseudorapidity. Each of the $h$ cells emits $m$ hadrons in 
the same universal manner, hence
\begin{eqnarray}
n_A= \sum_{i=1}^{h_A} m_i.  \label{eq:nA} 
\end{eqnarray}Then, in full analogy to Eq.~(\ref{eq:pstat}), we obtain
\begin{eqnarray}
\langle n_A \rangle &=&  \langle m \rangle \langle h_A \rangle, \nonumber \\
\br{\Delta n_A}^2 &=& {\rm var}(m) \langle h_A \rangle + \langle m \rangle^2 \br{\Delta h_A}^2,\nonumber  \\ 
\br{\Delta n_A \Delta n_{A'}} &=& \langle m \rangle^2 \br{\Delta h_A \Delta 
h_{A'}},\label{eq:hnstat}\\
\br{\Delta n^3_A} &=& \mu_3\bn{m}+3\var{m}\br{m}\br{\Delta h_A^2}\nonumber\\&+&\brp{m}{3}\br{\Delta h_A^3},\nonumber\\
\br{\Delta n_A \Delta n^2_{A'}} &=& \var{m}\br{m}\br{\Delta h_A \Delta 
h_{A'}}\nonumber\\&+&\brp{m}{3}\br{\Delta h_A \Delta h^2_{A'}},  \nonumber\\
\br{\Delta n_F \Delta n_C \Delta n_B} &=& \langle m \rangle^3 \br{\Delta h_F \Delta h_C \Delta h_B}.\nonumber
\end{eqnarray}
The numbers $n_A$ are observed experimentally. We note that the distribution of $m_i$ incorporates the effects of the 
detector acceptance. 

\subsection{Combined formulas}

We are now ready to combine 
Eqs.~(\ref{eq:pstat},\ref{eq:hstat},\ref{eq:hnstat}), which yield relations 
between moments of produced hadron distributions and moments of the initial sources: 
\begin{flushleft}
\begin{eqnarray}
        \br{n_A} &=&  \alpha\br{s_A},\nonumber\\
   \br{\Delta n_{A}^2} &=& \beta\br{s_A}+\gamma \br{\Delta s_A^2}\nonumber\\
         \br{\Delta n_{A} \Delta n_{A'}} &=&  \gamma \br{\Delta s_{A} \Delta 
s_{A'}},\label{eq:bigpack}\\
          \br{\Delta n_{A}^3} &=& \zeta\br{s_A}+3\gamma^{1/2}\beta' \br{\Delta s_A^2}+\gamma^{3/2}\br{\Delta s_{A}^3},\nonumber\\
          \br{\Delta n_{A} \Delta n_{A'}^2} &=&  \gamma^{1/2}\beta' \br{\Delta 
s_{A} \Delta s_{A'}}+\gamma^{3/2}\br{\Delta s_{A} \Delta s_{A'}^2},\nonumber\\
          \br{\Delta n_{F} \Delta n_{C} \Delta n_{B}} &=&  \gamma^{3/2}
\br{\Delta s_{F} \Delta s_{C} \Delta s_{B}}. \nonumber
       \end{eqnarray}
\end{flushleft}
The constants introduced above are defined as
\begin{eqnarray}
 \alpha &=& t_0\br{\mu}\br{m}, \nonumber \\ 
 \beta &=&  t_0 \br{\mu}\var{m}+t_1^2\brp{m}{2}\var{\mu},\label{eq:param}\\
 \beta' &=& t_1 \br{\mu}\var{m}+t_1^2\brp{m}{2}\var{\mu},\nonumber \\
 \gamma &=& t_1^2\br{m}^2\brp{\mu}{2},\nonumber\\
 \zeta &=& t_0 \mu_3(m)\br{\mu}+3 t_1^2\var{\mu}\var{m}\br{m}+t_1^3\mu_3(\mu)\brp{m}{3}.\nonumber
\end{eqnarray}
The algebraic structure of Eqs.~(\ref{eq:bigpack}) is very simple. However, nontrivial character stems from the fact that 
the parameters (\ref{eq:param}) do not depend on centrality of the collision.
        
It is convenient to introduce a short-hand notation
\begin{eqnarray}
  S_{ijk} = \br{\Delta s_B^{i} \Delta s_C^{j} \Delta s_F^{k} }
       \end{eqnarray}
for the moments of rank $r=i+j+k$, and
   \begin{eqnarray}
           \tilde{S}_{ijk} = \frac{\br{\Delta s_B^{i} \Delta s_C^{j} \Delta s_F^{k} }}{\sqrt{\br{\Delta s_B^{2}} \br{\Delta s_C^{2}} \br{\Delta s_F^{2}}}} \label{eq:momsc}
   \end{eqnarray}
for the moments scaled with the standard deviations. 
In particular, the forward-backward correlation of the number of sources is 
  \begin{eqnarray}
    \tilde{S}_{101}&\equiv&\frac{\br{\Delta s_B \Delta s_F }}{\sqrt{\br{\Delta s_B^{2}} \br{\Delta s_F^{2}}}}\equiv \rho(s_F,s_B).\label{eq:fbinit}
  \end{eqnarray}
The scaled variance, used in the following, is defined as 
\begin{eqnarray}
\omega(x_A) = {\rm var}(x_A)/\langle x_A  \rangle. 
\end{eqnarray}
We also introduce the combinations of parameters 
\begin{eqnarray}
\delta &=& \beta/\alpha = \omega(m) +\frac{t_1^2}{t_0} \langle m \rangle \omega(\mu), \nonumber \\
\kappa &=& \gamma/ \alpha = \frac{t_1^2}{t_0} \langle \mu \rangle \langle m \rangle, \label{eq:kap} \\
\lambda &=& \beta/\gamma= \frac{t_0 \omega(m)}{t_1^2 \langle \mu \rangle\langle m \rangle} + \frac{\omega(\mu)}{\langle \mu \rangle}. \nonumber 
\end{eqnarray}

\section{General results \label{sec:results}}

\subsection{Relations of moments of hadrons to moments of sources }

Our methodology uses the Glauber model for the sources to make predictions for the moments of $n_A$.

An interesting relation for the hadron correlations results from Eq.~(\ref{eq:bigpack}):
\begin{eqnarray}
\rho(n_A,n_{A'}) = {\rho(s_A,s_{A'})}{\sqrt{\left({1-\frac{\delta}{\omega(n_A)}}\right )\left( {1-\frac{\delta}{\omega(n_{A'})}}\right ) }}.
\nonumber \\ \label{eq:omr0}
\end{eqnarray}
Since, as we shall see, for not too large pseudorapidity separations, in the Glauber model to a high accuracy $\rho(s_A,s_{A'})\simeq 1$, it then follows that 
\begin{eqnarray}
\rho(n_A,n_{A'}) \simeq {\sqrt{\left ( {1-\frac{\delta}{\omega(n_A)}}\right ) \left ( {1-\frac{\delta}{\omega(n_{A'})}}\right )}}. \label{eq:omr1}
\end{eqnarray}
which means a very specific relation between the correlation of numbers of hadrons in bins $A$ and $A'$ and the corresponding scaled variances, 
which depends on a single {\em centrality independent} parameter $\delta$. Relation (\ref{eq:omr1}) may be straightforwardly tested with the future data. 

From Eqs.~(\ref{eq:bigpack}) we also find
\begin{eqnarray}
\omega(n_A)   &=& \delta + \kappa \omega(s_A),  \nonumber \\
\rho(n_F,n_B) &\equiv& \frac{\rho(s_F,s_B)}{\sqrt{\bn{1+\frac{\lambda}{\omega(s_F)}}\bn{1+\frac{\lambda}{\omega(s_B)}}}}, \label{eq:omr}
\end{eqnarray}
formulas presented already in Ref.~\cite{Olszewski:2013qwa}. 

Another interesting relation, linking three bins, has the form
\begin{eqnarray}
\frac{\br{\Delta n_B \Delta n_C \Delta n_F}}{\br{\Delta n_A \Delta n_{A'}}^{3/2}}=\frac{\br{\Delta s_B \Delta s_C \Delta s_F}}{\br{\Delta s_A \Delta s_{A'}}^{3/2}}\label{eq:ratio}.
\end{eqnarray}
Again, the right-hand side, obtainable in a model, can be verified against the future data, yielding the left-hand side.

With the help of the basic formulas~(\ref{eq:bigpack}) it is straightforward to find relations for the newly-proposed 
measures~\cite{Gorenstein:2011vq,Gazdzicki:2013ana} of fluctuations. 
They are defined as 
\begin{eqnarray}
\Sigma(n_F,n_B)&=&\frac{\langle n_F \rangle \omega(n_B) + \langle n_B \rangle \omega(n_F) - 2{\rm cov} (n_F,n_B)}{C_\Sigma}, \nonumber \\
\Delta(n_F,n_B)&=&\frac{ \langle n_B \rangle \omega(n_F) - \langle n_F \rangle \omega(n_B)}{C_\Delta},  \label{eq:side}
\end{eqnarray}
where the constants $C_\Sigma$ and $C_\Delta$ are proportional to first moments of any extensive quantity.  We take the simplest choice 
\begin{eqnarray}
C_\Sigma = C_\Delta = \langle n_F \rangle + \langle n_B \rangle.
\end{eqnarray}
Elementary algebra leads to the relations
\begin{eqnarray}
\Sigma(n_F,n_B)&=&\delta + \kappa \Sigma(s_F,s_B), \nonumber \\
\Delta(n_F,n_B)&=&\delta \frac{\langle s_F-s_B \rangle}{\langle s_F+s_B \rangle} + \kappa \Delta(s_F,s_B), \label{eq:side1}
\end{eqnarray}
where $\Sigma(n_F,n_B)$ and $\Delta(n_F,n_B)$ are defined analogously to Eq.~(\ref{eq:side}), but with $n_A$ replaced with $s_A$.

\subsection{Implications for the qualitative features of hydrodynamics}

\begin{figure}
\begin{center}
\includegraphics[scale=0.65]{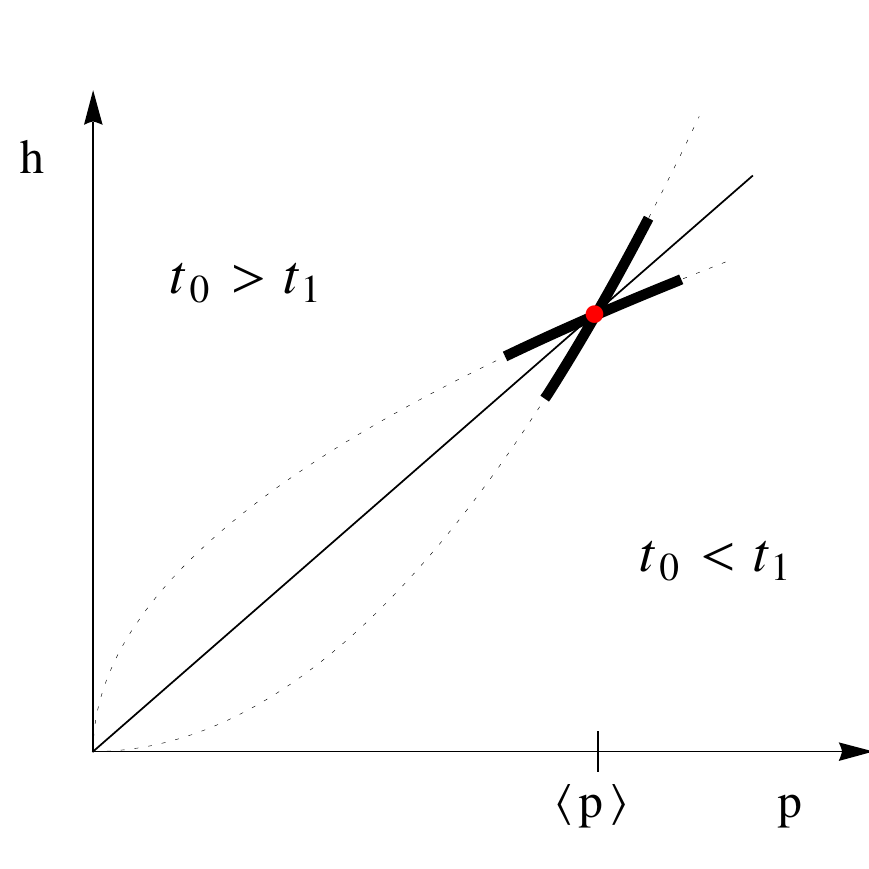}
\caption{(color online) Possible scenarios for the hydrodynamic response (see text for details).
\label{fig:hydro}}
\end{center}
\end{figure}

Based on first and third of Eqs.~(\ref{eq:bigpack}), we find the relation
 \begin{eqnarray}
\frac{t_0}{t_1}=\frac{\alpha}{\gamma^{1/2}}. \label{eq:rrr}
 \end{eqnarray}
Since the $\alpha$ and $\gamma$ parameters may be found by fitting the model predictions to the future data, we are able 
to obtain from Eq.~(\ref{eq:rrr}) information concerning the hydrodynamic response in the intermediate phase. 
Figure~(\ref{fig:hydro}) shows two possible scenarios for the hydro 
evolution. When
$t_0 > t_1$ ($t_0 < t_1$), the hydrodynamic growth $h(p)$ is slower (faster) than the linear 
function. From the experience of hydrodynamics we expect that $t_0 \simeq t_1$, which means that the entropy in the 
considered cell is proportional to its initial entropy  (see,  e.g.,~\cite{Bozek:2011ua}). 

\section{Dependence of results on width of centrality bins \label{sec:cen}}

An important question concerns the  dependence of the 
results shown in this paper on centrality selection. 
In our analysis presented below we define centrality with the help of the number of sources in 
the central bin, $s_C$.
The results for moments of multiplicities depend on 
the chosen width of the centrality bin $\Delta c$ -- in that sense they are not 
{\em intensive measures} \cite{Gorenstein:2011vq,Mrowczynski:1999un}. Nevertheless, the 
moments evaluated that way are still useful for comparing model predictions to experiment, as exactly the 
same choice of width of the centrality windows can be made in model simulations 
as in the data analysis. 

\begin{figure}
\begin{flushleft}
\includegraphics[scale=.9]{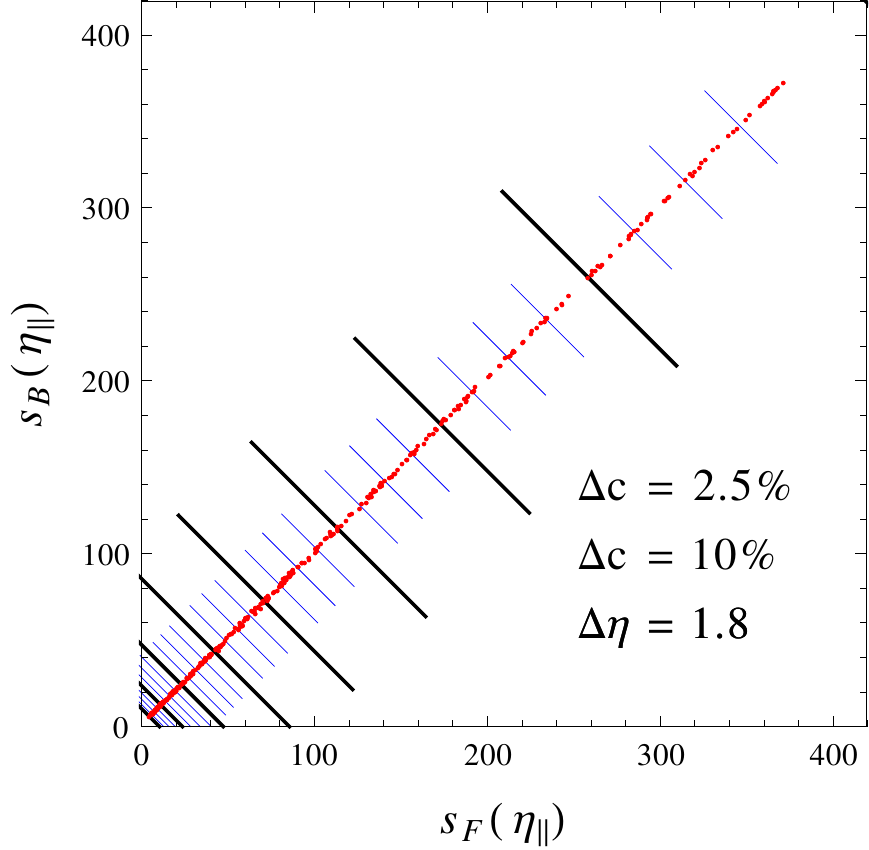}
\caption{(color online) Scattered plot for the forward-backward correlation of the numbers of initial sources $s_F$ and $s_B$ for rapidity 
separation $\Delta\eta=1.8$. The short lines cut the sample into centrality bins of width $\Delta c = 2.5\%$, and the long lines into bins of width $\Delta c = 10\%$.
{\tt GLISSANDO}, mixed model, Pb+Pb collisions at the LHC energy of $\sqrt{s_{NN}}=2.76 \textrm{ TeV}$.
\label{fig:scater}}
\end{flushleft}
\end{figure}

To understand the problem, let us focus on  Fig.~\ref{fig:scater}, where we show a result of a simulation
in the form of a scattered plot of $s_F$ and $s_B$. The short skewed lines cut the events into 
subsequent centrality classes with \mbox{$\Delta c=2.5\%$}, while the longer lines 
into classes with $\Delta c=10\%$ (containing 4 bins of $\Delta c=2.5\%$). We note that 
the scattered plot is very much elongated, simply reflecting the large correlation between
$s_F$ and $s_B$. With the exception of very small centralities $c$ (the lower left 
corner of the plot), the correlation between $s_F$ and $s_B$ within a given 
centrality window is very close to 1, $\rho(s_F,s_B)\simeq 1$. As we move 
towards very low $c$ (lower-left corner), or when $\Delta c$ is decreased, $\rho(s_F,s_B)$ becomes 
lower than 1. Thus the correlation of the sources is in an obvious way affected by the choice of the centrality bins. 

For sufficiently large width of the bins, the above-discussed feature allows us for a derivation  of 
approximate scaling laws of the moments on $\Delta c$. 
Details are shown in Appendix~\ref{sec:cenexp}. The 
average multiplicities do not scale, as long as the window is sufficiently 
narrow:
\begin{eqnarray}
\langle s_A \rangle \sim (\Delta c)^0.  \label{eq:nsr0}
\end{eqnarray}
For the variance and covariance we find
\begin{eqnarray}
\langle (\Delta s_A)^2 \rangle \sim (\Delta c)^2, \nonumber \\
\langle \Delta s_A \Delta s_{A'} \rangle \sim (\Delta c)^2, \label{eq:nsr}
\end{eqnarray}
from where 
\begin{eqnarray}
\rho(s_A,s_{A'}) \sim (\Delta c)^0.
\end{eqnarray}
For the moments of rank 3 we need to be a bit more sophisticated.
The marginal 
distribution for the variable $s_A$ in a given centrality bin can be 
approximated as an affine function, $f(s_A) = N(1+a s_A)$, where $a$ is small. 
Then (cf. Appendix~\ref{sec:cenexp}), to leading order in $a$
\begin{eqnarray}
\langle (\Delta s_A)^3 \rangle \sim a (\Delta c)^4,
\end{eqnarray}
and 
\begin{eqnarray}
\langle (\Delta s_A)^3 \rangle / \langle \Delta s_A \Delta s_{A'} 
\rangle^{3/2} \sim a \Delta c. \label{eq:adv}
\end{eqnarray}
The above scaling laws are visible in our results presented below.   

\section{{\tt GLISSANDO} simulations \label{sec:glis}}

We use GLISSANDO~\cite{Rybczynski:2013yba} to model the initial phase.  To simulate Pb+Pb collisions at the LHC at 
\mbox{$\sqrt{s_{NN}}=2.76 \textrm{ TeV}$}, we use (unless stated otherwise) the mixed Glauber 
model ~\cite{Kharzeev:2000ph,SchaffnerBielich:2001qj,Back:2001xy} with the mixing parameter \mbox{$a=0.1$} (cf. Appendix~\ref{sec:emit}),
the NN cross section $\sigma_{NN}=66.3 \textrm{ mb}$ and a 
Gaussian wounding profile~\cite{Rybczynski:2011wv}. The numbers of sources $s_A$ are obtained in pseudo-rapidity bins of width $\delta\eta=0.2$, while 
$\Delta \eta$ denotes the separation of the centers of the $F$ and $B$ bins. The $C$ bin is centered around zero, while the $F$ and $B$ bins are 
arranged symmetrically.

\begin{figure}
\begin{flushleft}
\includegraphics[scale=.87]{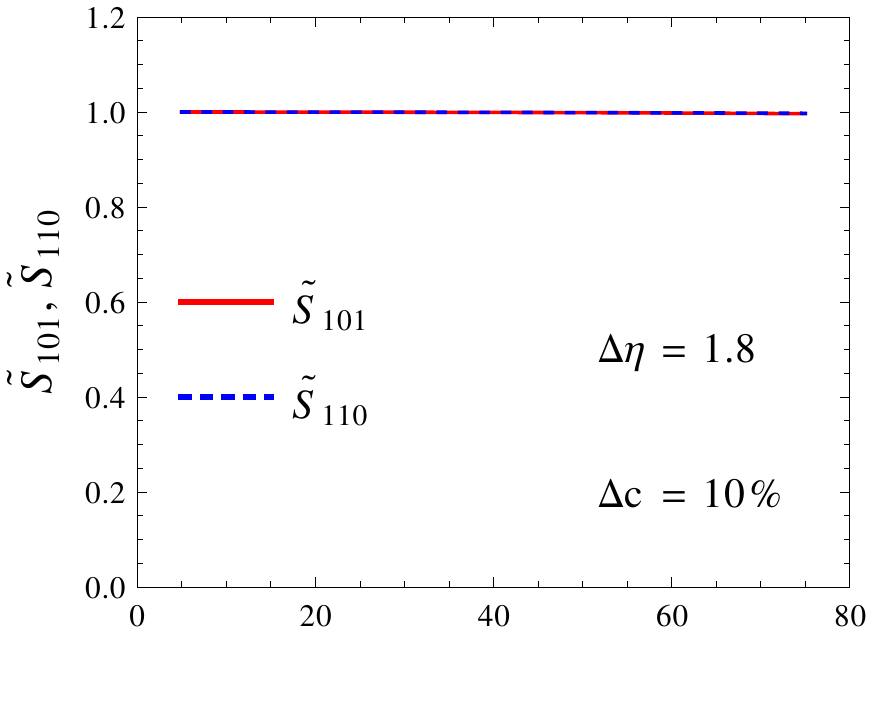}
\includegraphics[scale=.87]{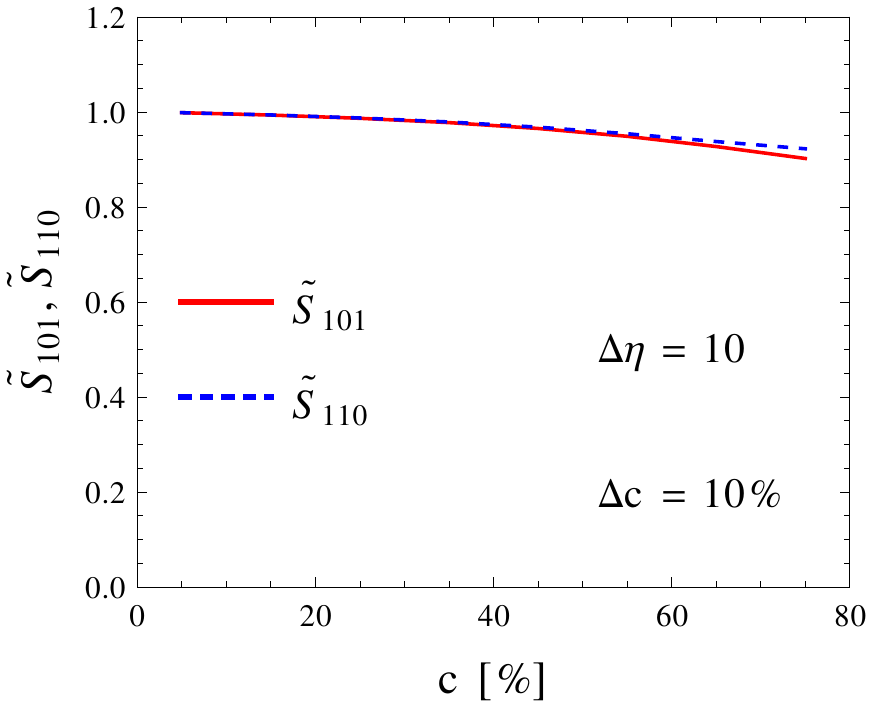}
\caption{(color online) Forward-backward and forward-central correlations of the number of initial sources, plotted as functions of centrality
for two cases of the pseudorapidity separation, $\Delta\eta=1.8$ and $\Delta\eta=10$. 
\label{fig:fbinit}}
\end{flushleft}
\end{figure}

\begin{figure}
\begin{flushleft}
\includegraphics[scale=.87]{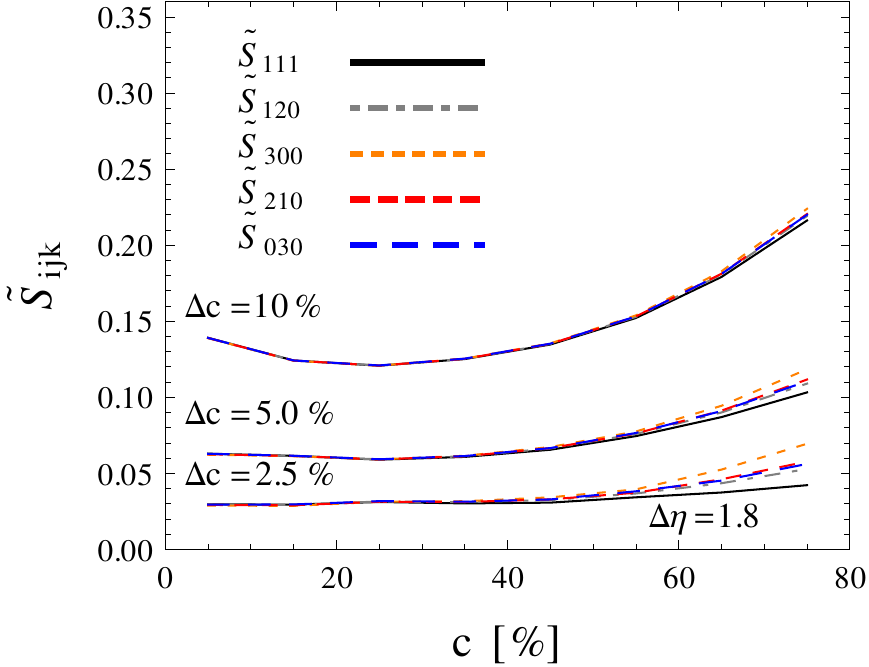}
\caption{(color online) Scaled rank-3 moments of sources plotted as functions of centrality for three sample values of the width for the centrality bins. 
\label{fig:moments}}
\end{flushleft}
\end{figure}

\subsection{Results for the correlations of sources}

We begin presenting our results with the forward-backward and forward-central correlation functions, $ \tilde{S}_{101}$ and  $\tilde{S}_{110}$, shown in Fig.~\ref{fig:fbinit} for the rapidity separations 
$\Delta \eta =1.8$ and  $\Delta \eta =10$. We use $\Delta c=10$~\%, but as follows from Eq.~(\ref{eq:nsr}), the results are insensitive to the width of the bin. We note that 
these correlation functions are essentially equal to unity, except for very large values of $\Delta \eta$ and $c$. This is an interesting feature of the adopted Glauber model, as it 
shows that we can use $\rho(s_A,s_{A'})\simeq 1$ in practical applications. This fact has its origin in a strong correlation between the numbers of wounded nucleons from 
nuclei A and B, as well as between the wounded nucleons and the binary collisions (cf. Appendix~\ref{sec:emit}).

For the case of symmetric collisions and symmetrically placed F and B bins, the measures $\Sigma(s_F,s_B)$ and $\Delta(s_F,s_B)$ from Eq.~(\ref{eq:side}) 
are equal to 
\begin{eqnarray}
\Sigma(s_F,s_B)&=&\omega(s_A)[1-\rho(s_F,s_B)], \label{eq:side3} \\
\Delta(s_F,s_B)&=&0, \hspace{3cm} (A=F,B), \nonumber
\end{eqnarray}
Since $\rho(s_F,s_B)\simeq 1$, it follows that $\Sigma(s_F,s_B) \simeq 0$.

The scaled rank-3 correlations of Eq.~(\ref{eq:momsc}) are displayed in Fig.~\ref{fig:moments}. We note the advocated scaling with the value of $\Delta c$, 
cf. Eq.~(\ref{eq:adv}). Moreover, various rank-3 moments shown in the plot are essentially equal to one another in the Glauber treatment. 

\begin{figure}
\begin{flushleft}
\includegraphics[scale=.87]{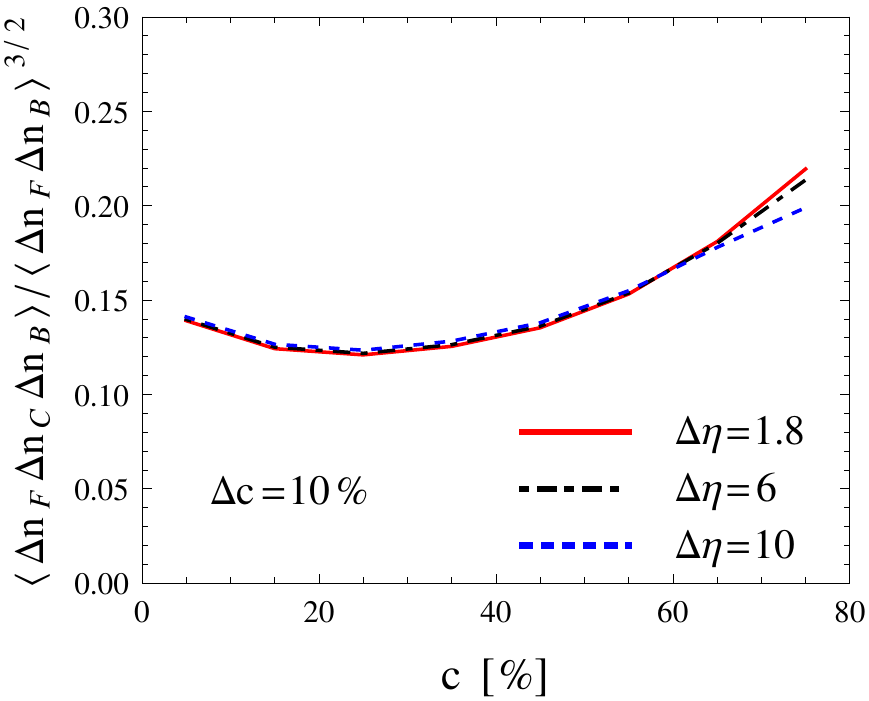}
\caption{(color online) Prediction for the combination of  third and second moments of observed hadrons, 
plotted as a function of centrality for three pseudo-rapidity separations, $\Delta\eta=1.8,\ 6,\ 10$.
\label{fig:predrat}}
\end{flushleft}
\end{figure}

\subsection{Results for the correlations of produced hadrons}

\begin{figure}
\begin{flushleft}
\includegraphics[scale=.87]{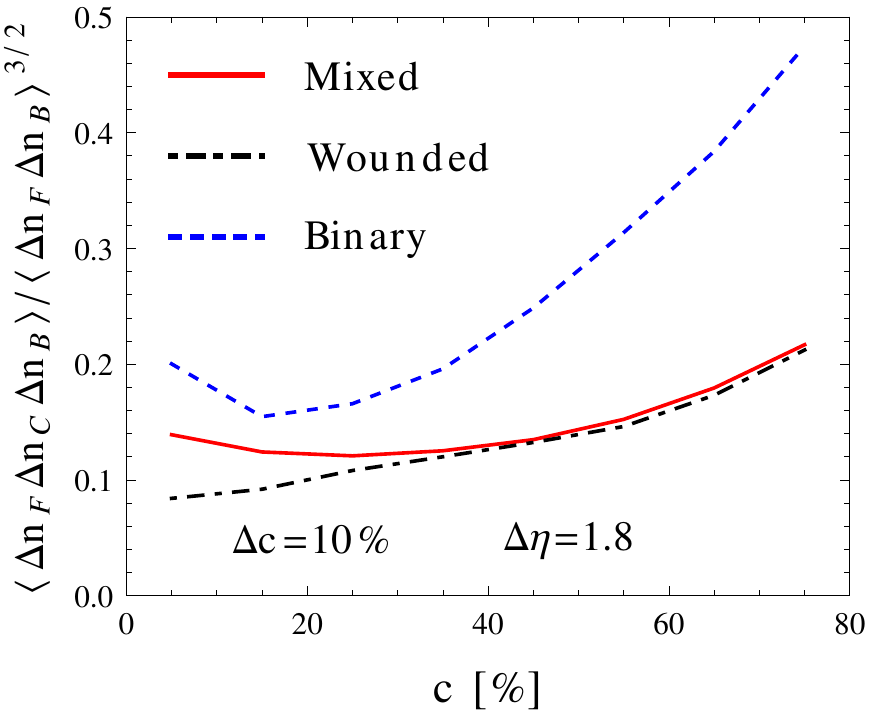}
\caption{(color online) Same as in Fig.~\ref{fig:predrat}, but for different variants of the Glauber model.
\label{fig:predrat1}}
\end{flushleft}
\end{figure}

Now we pass to the presentation of results for the moments of the distributions of the produced hadrons, $n_A$. Under the assumptions of the superposition 
approach, Eq.~(\ref{eq:ratio})
allows us to make predictions for the produced hadrons, based on the model of sources. Our result, for several pseudo-rapidity separations $\Delta \eta$, 
is shown in Fig.~\ref{fig:predrat}. The result is specific to the adopted source production model. 
In Fig.~\ref{fig:predrat1} we compare the outcome, for the same quantity,  of the wounded nucleon model 
(mixing parameter $a=0$), the mixed model ($a=0.1$), and the binary-collisions model ($a=1$). 
The large differences between the models should allow, with the future data, to discriminate between them.

\begin{figure}
\begin{flushleft}
\includegraphics[scale=.87]{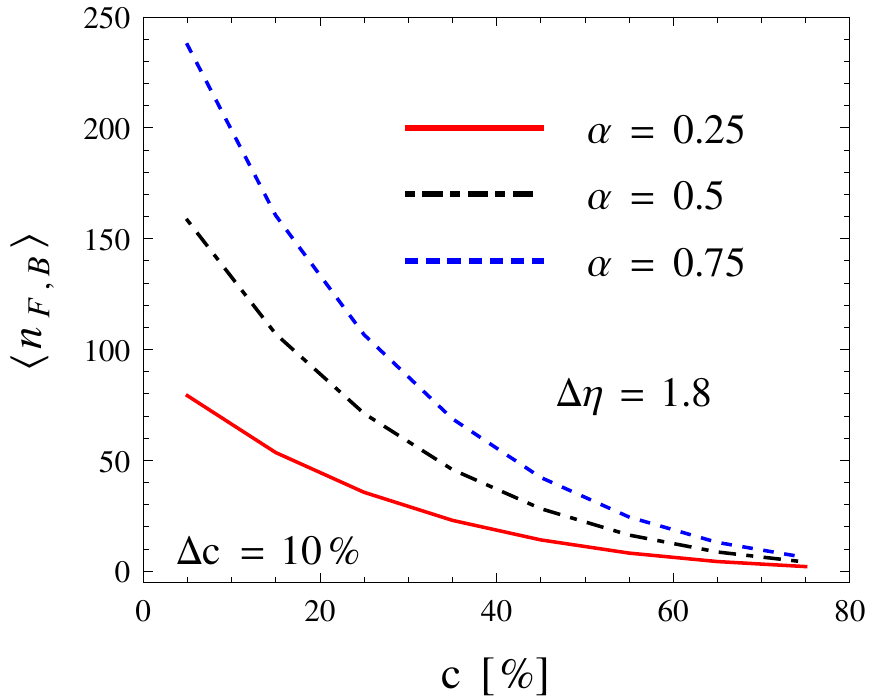}
\caption{(color online) Predictions for the multiplicity of the produced hadrons in peripheral bins, plotted as a functions of centrality, based on Eq. (\ref{eq:bigpack})
at different values of parameters in Eq.~(\ref{eq:param},\ref{eq:kap}).
\label{fig:mult}}
\end{flushleft}
\end{figure}

\begin{figure}
\begin{flushleft}
\includegraphics[scale=.87]{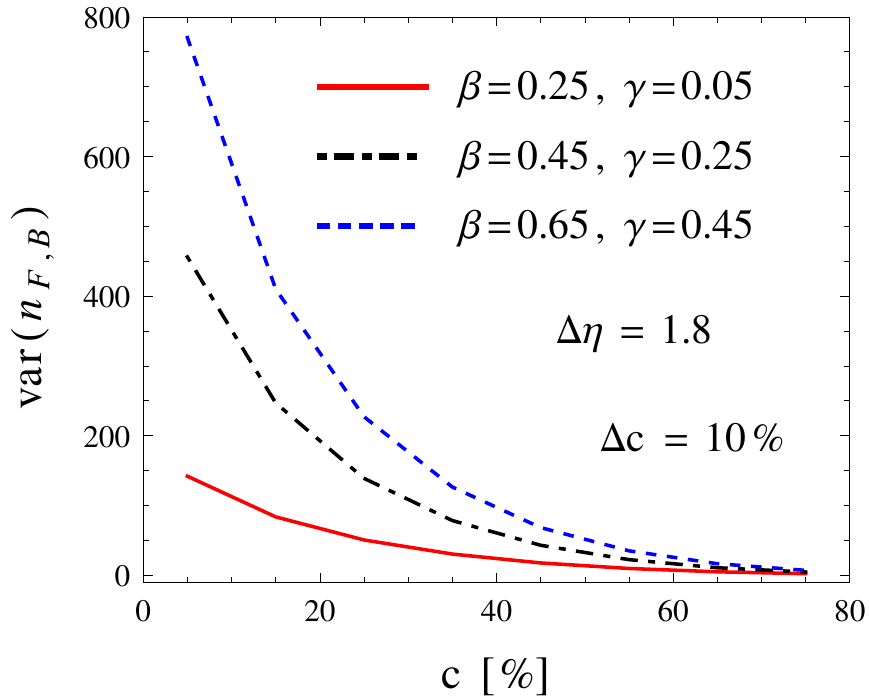}
\caption{(color online) Same as Fig.~\ref{fig:mult}, but for the variance.
\label{fig:var}}
\end{flushleft}
\end{figure}

\begin{figure}
\begin{flushleft}
\includegraphics[scale=.87]{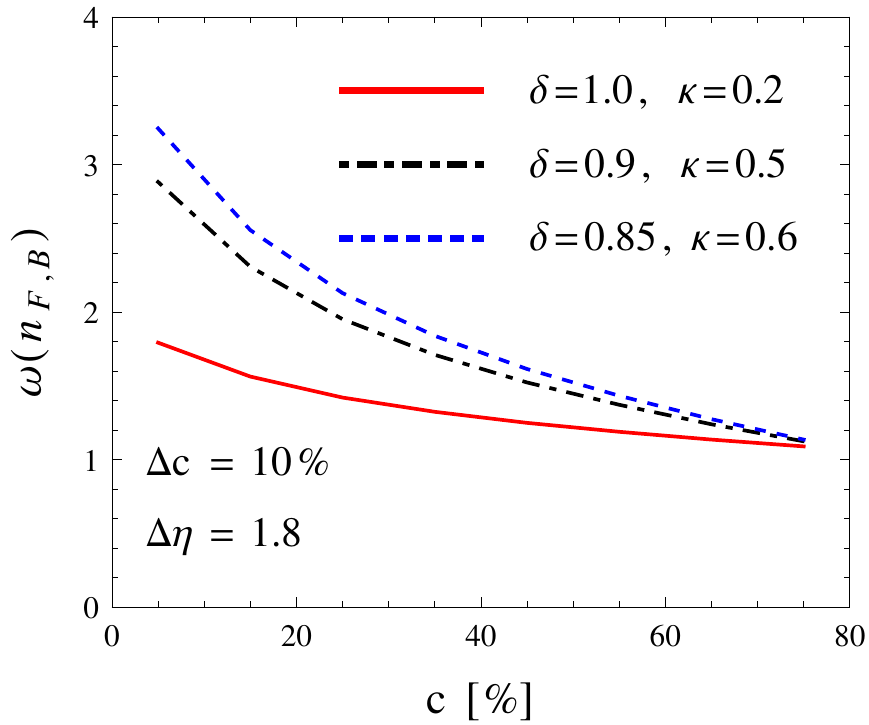}
\caption{(color online) Same as Fig.~\ref{fig:mult}, but for the scaled variance.
\label{fig:omega}}
\end{flushleft}
\end{figure}

\begin{figure}
\begin{flushleft}
\includegraphics[scale=.87]{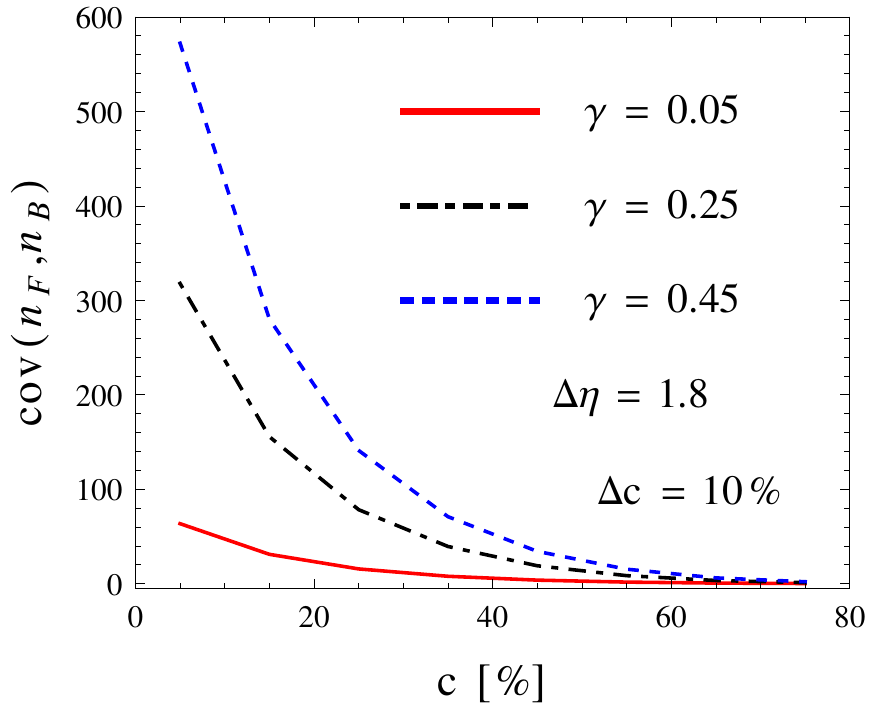}
\caption{(color online) Same as Fig.~\ref{fig:mult}, but for the covariance.
\label{fig:cov}}
\end{flushleft}
\end{figure}

\begin{figure}
\begin{flushleft}
\includegraphics[scale=.87]{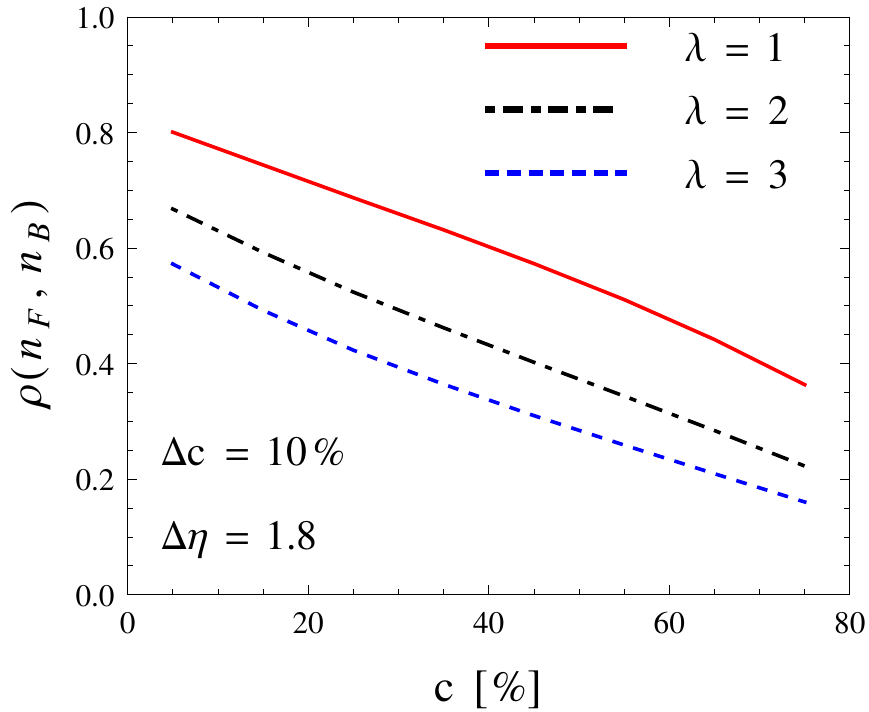}
\caption{(color online) Same as Fig.~\ref{fig:mult}, but for the forward-backward correlations.
\label{fig:predfb}}
\end{flushleft}
\end{figure}

Finally, we use Eq. (\ref{eq:bigpack}) with some reasonable choice of the parameters (\ref{eq:param},\ref{eq:kap}) 
to make predictions for the multiplicity, variance, covariance, and rank-2 correlation coefficients. The results are displayed in Figs.~\ref{fig:mult}-\ref{fig:predfb}.
We may now carefully illustrate our methodology: with the data for multiplicity we fit the value of $\alpha$ with the results of Fig.~\ref{fig:mult},
the value of $\gamma$ with Fig.~\ref{fig:cov}, and then the value of $\beta$ with Fig.~\ref{fig:var}. Next, we evaluate  $\lambda=\beta/\gamma$ and make a test 
with the help of Fig.~\ref{fig:predfb}. Extension to rank-3 moments is straightforward. That way, we may obtain the parameters of the superposition model in a systematic way.
Knowledge of these parameters will provide insight into the microscopic features of the particle production mechanism in the system.

\subsection{Centrality from the number of hadrons \label{sec:cn}}

In this Section we check how the results for the forward-backward correlations depend on the method of determining centrality, carrying out a study similar 
to Ref.~\cite{Konchakovski:2008cf}. In Sec.~\ref{sec:cendef} we have introduced the definition of centrality, used 
in the remaining parts of this paper, based on the number of 
sources in the central rapidity bin, $s_C$. To estimate the results based on centrality determined with number of produced hadrons in the central bin, $n_C$, we carry 
out {\tt GLISSANDO} simulations, where a negative binomial distribution is superimposed over the distribution of sources. The parameters of the binomial distributions are such that the 
experimental multiplicity distributions at the LHC are reproduced~\cite{Bozek:2015bna}.

The qualitative effect of the superimposed 
fluctuations is simple to understand: when the limits of the centrality bin are $n_C^{\rm low}$ and  $n_C^{\rm high}$, the corresponding limits in  the 
number of sources, $s_C^{\rm low}$ and  $s_C^{\rm high}$ are somewhat diffused. This causes more extended scattered plot of type of Fig.~\ref{fig:scater} for the events belonging to 
the considered centrality class and the correlation coefficient $\rho(s_F,s_B)$ increases. As may be inferred form Fig.~\ref{fig:scater}, the effect will be larger for peripheral collisions, where 
the scattered plots are less elongated, and smaller for central collisions, where elongation (for $\Delta c=10\%)$ is
large.

\begin{figure}
\begin{flushleft}
\includegraphics[scale=1.28]{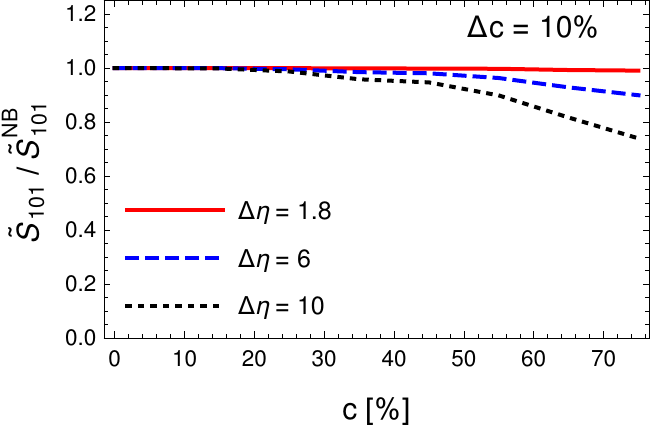}
\caption{(color online) Influence of the method of centrality selection for the forward-backward correlation of sources, tested with the ratio $\tilde{S}_{101}/\tilde{S}^{\rm NB}_{101}$
of the forward-backward correlations evaluated with the number of sources $s_C$ and the number of hadron $n_C$ in the central rapidity bin. 
At small forward-backward 
separation, $\Delta \eta =1.8$, there is no effect of the method used  to determine centrality. 
\label{fig:rhocen}}
\end{flushleft}
\end{figure}

Results of our numerical simulations are presented in Fig.~\ref{fig:rhocen}, where we show the ratio of the correlation coefficients $\tilde{S}_{101}$ for the 
centralities evaluated according to $s_C$ and according to $n_C$ obtained with the overlaid negative binomial distribution, indicated as $\tilde{S}_{101}/\tilde{S}^{\rm NB}_{101}$. 
We note that for small forward-backward pseudorapidity separation, $\Delta \eta=1.8$, there is no visible effect, whence  $\tilde{S}_{101}/\tilde{S}^{\rm NB}_{101} \simeq 1$. 
For large  $\Delta \eta$ and peripheral collisions $\tilde{S}_{101}/\tilde{S}^{\rm NB}_{101} < 1$, in agreement with the qualitative arguments presented above.

\section{Conclusions \label{sec:cls}}

We have presented a systematic study of two- and three-rapidity-bin multiplicity correlations in relativistic heavy-ion 
collisions in a superposition approach, based on three phases of the collision: initial partonic phase, intermediate 
collective evolution, and final statistical hadronization. 
The derived expressions show how the moments of 
the produced particles are related to the moments of the sources produced in the initial phase. The relations, simple but 
nontrivial, involve only a few parameters collecting the microscopic information concerning the dynamics (parameters of the 
superimposed distributions, response of hydrodynamics).

With a definite model for the production of sources, such as the Glauber model used here, one is then able to obtain 
in a systematic way
predictions for the multibin multiplicity correlations of the produced hadrons. Conversely, with the future experimental data, one will be able to 
test the models of the initial phase, as well as obtain information on the microscopic parameters. 
In particular, the awaited forward-backward multiplicity correlation analysis with the LHC data will shed light on the early production mechanism. 
The statistical method presented in this paper is directly applicable to that case. 

\begin{acknowledgments}
This research was supported by the Polish National Science Centre
grants DEC-2012/05/B/ST2/02528 and DEC-2012/06/A/ST2/00390.
\end{acknowledgments}

\appendix

\section{Centrality width dependence\label{sec:cenexp}}

In this Appendix we sketch the derivation of scaling of the moments with the width of the centrality bin.
The explanation comes from Fig. (\ref{fig:scater}), where the forward-backward correlation is plotted as scattered plot. 
For a given centrality and for a typical size of $\Delta c$ we may approximate the distribution of $s_A=x$ as an affine 
function,
\begin{eqnarray}
 f(x)&=&\frac{1+a x}{\bn{1+a c}\Delta c},
 \end{eqnarray}
 which has been normalize to 1. Upon expanding for small $a c$ it follows that 
 \begin{eqnarray}
 && \br{x}=\int_{c-\frac{\Delta c}{2}}^{c+\frac{\Delta c}{2}}x f(x)\simeq c+\frac{a \Delta c^2}{12}, \\
 && \br{\bn{\Delta x}^2}=\int_{c-\frac{\Delta c}{2}}^{c+\frac{\Delta c}{2}}\bn{x-\br{x}}^2\simeq \frac{ \Delta c^2}{12}, \nonumber \\
 && \br{\bn{\Delta x}^3}=\int_{c-\frac{\Delta c}{2}}^{c+\frac{\Delta c}{2}}\bn{x-\br{x}}^3\sim \Delta c^4, \nonumber \\
 && \frac{\br{\bn{\Delta x}^3}}{\br{\bn{\Delta x}^2}^{3/2}}\sim  \Delta c. \nonumber
\end{eqnarray}
The scaling laws (\ref{eq:nsr0},\ref{eq:nsr}) result from the above formulas.

\section{Emission profiles in spatial rapidity \label{sec:emit}}

\begin{figure*}
\begin{center}
\includegraphics[scale=.87]{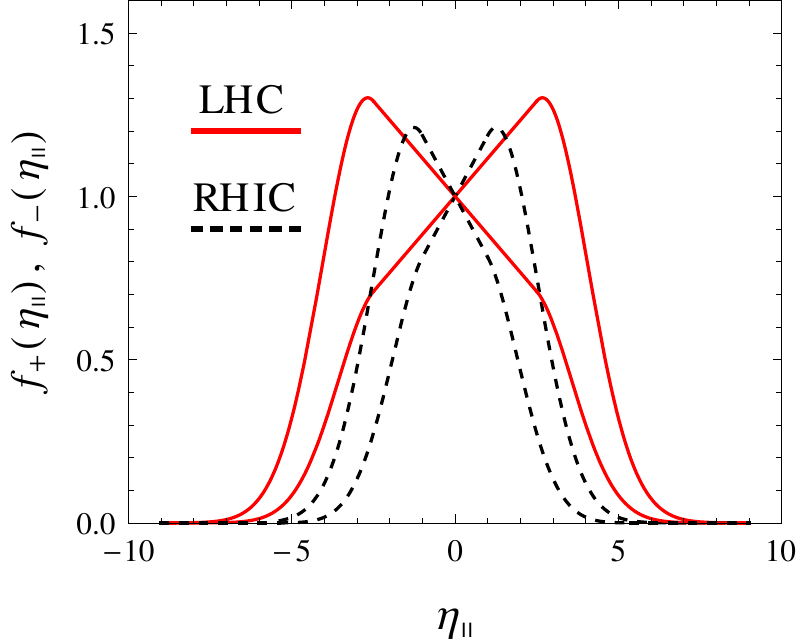} \includegraphics[scale=.87]{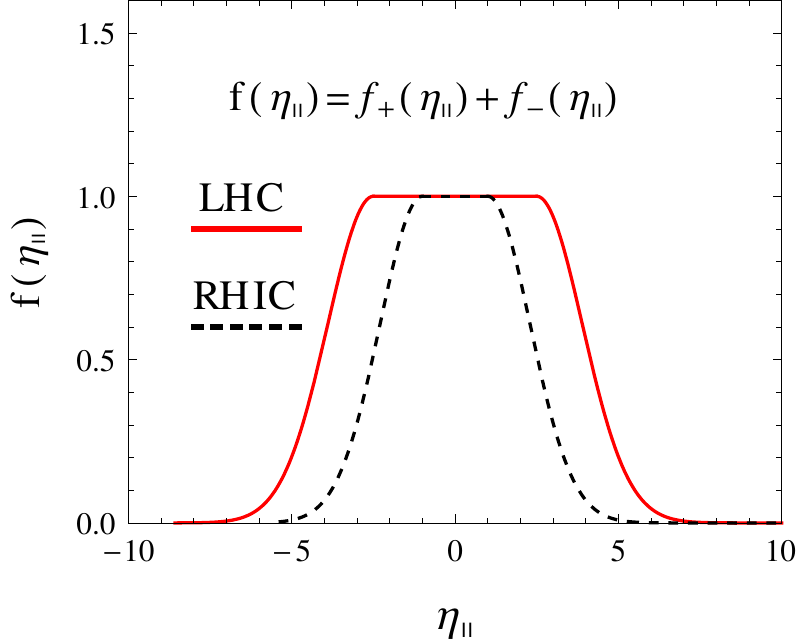}
\caption{(color online) The emission profiles in space-time rapidity for the wounded nucleons (dashed lines) and the binary 
collisions (solid line). The profile $f_+ (f_-)$ corresponds to the forward (backward) moving wounded nucleons.
\label{fig:prof}}
\end{center}
\end{figure*}

In a nucleus-nucleus collision, the emission profile~\cite{Bozek:2010bi} defining the shape of initial sources 
in the spatial rapidity has the form
\begin{eqnarray}
s\!\bn{\eta_{||}}&=&\bn{\frac{1-a}{2}}\bn{f_+\!\bn{\eta_{||}} N_{w,B}+f_-\!\bn{\eta_{||}} N_{w,F}}\nonumber\\
&+&a \bn{\frac{f_+\!\bn{\eta_{||}}+f_-\!\bn{\eta_{||}}}{2}} N_{bin},\label{eq:source}
\end{eqnarray}
where $N_{w,F}$ and $N_{w,F}$ are the numbers of the forward- and backward-going wounded nucleons, $N_{bin}$ is the number of binary collisions,
while $f_{+}\bn{\eta_{||}}$ and $f_{-}\bn{\eta_{||}}$ describe the corresponding wounded-nucleon emission profiles. 
Finally, $f\bn{\eta_{||}}$ is the emission profile for the binary collisions.
We use the following parametrization~\cite{Bozek:2010bi}:
    \begin{eqnarray}
     f\bn{\eta_{||}}&=&\exp\!\bn{-\frac{\bn{\vert \eta_{||}\vert - \eta_0}^2}{2 \sigma^2}\theta\!\bn{\vert \eta_{||}\vert - \eta_0}},\nonumber\\
     f_+\bn{\eta_{||}}&=&f_F\bn{\eta_{||}} f\bn{\eta_{||}}\label{eq:prof},\\
     f_-\bn{\eta_{||}}&=&f_F\bn{-\eta_{||}} f\bn{\eta_{||}},\nonumber
    \end{eqnarray}
with
\begin{displaymath}
f_F\bn{\eta_{||}} = \left\{ \begin{array}{ll}
0, & \eta_{||}\le -\eta_m\\
\frac{\eta_{||}+\eta_m}{2\eta_m} & -\eta_m < \eta_{||} < \eta_m\\
1 & \eta_m \le \eta_{||}.
\end{array} \right.
\end{displaymath}
Sample values of parameters~\cite{Bozek:2013uha}, describing the ATLAS data after the 
hydrodynamic evolution, are $\eta_0=1.25$, $\eta_m=8.58$, $\sigma_{\eta}=1.4$.
The functions (\ref{eq:prof}) are shown in Fig.~\ref{fig:prof}.

\bibliographystyle{apsrev4-1-nohep}
\bibliography{fromhep,hydr}

\end{document}